\documentclass[twocolumn,aps,apl,showpacs,showkeys]{revtex4} 

\usepackage{graphicx}
\usepackage{bm}
\usepackage{times}

\begin{document}


\title{Effects of magnetic field on two-dimensional Superconducting Quantum Interference Filters}

\author{J.~Oppenl{\"a}nder}%
\author{P.~Caputo}%
\email{p.caputo@uni-tuebingen.de}%
\author{Ch.~H{\"a}ussler}%
\author{T.~Tr\"auble}%
\author{J.~Tomes}%
\author{A.~Friesch}%
\author{N.~Schopohl}%
\affiliation{%
Lehrstuhl f\"ur Theoretische Festk{\"o}rperphysik, Universit{\"a}t T{\"u}bingen\\ 
Auf der Morgenstelle 14, 72076 T{\"u}bingen (Germany)
}%

\date{\today}

\begin{abstract}

We present an experimental study of two-dimensional superconducting quantum interference filters (2D-SQIFs) in the presence of a magnetic field $B$. The dependences of the dc voltage on the applied magnetic field are characterized by a unique delta-like dip at $B=0$, which depends on the distribution of the areas of the individual loops, and on the bias current. The voltage span of the dip scales proportional to the number of rows simultaneously operating at the same working point. In addition, the voltage response of the 2D-SQIF is sensitive to a field gradient generated by a control line and superimposed to the homogeneous field coil. This feature opens the possibility to use 2D superconducting quantum interference filters as highly sensitive detectors of spatial gradients of magnetic field.

\end{abstract}

\pacs{74.50.+r, 74.81.Fa, 85.25.Dq, 85.25.Cp}
\keywords{SQIF, arrays, Josephson junctions, magnetometry}

\maketitle


Superconducting Quantum Interference Filters (SQIFs) are one or two dimensional arrays of Josephson junctions (JJs) with a specially selected distribution of the loop areas%
\cite{Oppenlander:PRB,Haussler:JAP2001}. While in regular arrays with equal areas of the loops the dependence of the critical current on magnetic field $I^{ar}_{c}(B)$ (or voltage on magnetic field $V(B)$, at bias current $I_b>I^{ar}_{c}$) looks like periodically repeated Fraunhofer-like fragments, in SQIFs the corresponding dependence has only one main peak (or dip in case of $V(B)$) at $B=0$, with rather steep slopes. This property makes SQIFs very attracting for applications. SQIFs can be realized in a relative simple way by using well-established fabrication processes, both with low- \cite{Oppenlander:PhysicaC-2002,Oppenlander:ASC2001} and high-$T_c$ JJs \cite{Schultze:ASC2002}. Indeed, since the principle of operation of SQIFs is based on quantum interference of all cells, SQIFs are highly robust against the spread of the junction parameters, as it was shown theoretically and experimentally \cite{Oppenlander:PRB,Schultze:ASC2002}.

A typical 1D-SQIF studied by us earlier consists of a parallel array of $n\,=\,1,...,N$ Josephson junctions, with an unconventional distribution of the cell areas ${\bf a}_n$ \cite{Oppenlander:ASC2001}. 
The $V({\bf B})$ dependence is defined by the equation          
\[
  V(\mathbf{B}) = I_c R \sqrt{J_N^2-{|{S_N(\mathbf{B})}|}^2},
\]
where:
\[
  \left\{
  \begin{array}{lcl}
    I_c = \frac{1}{N}\sum\limits _{n=1}^{N} I_{c,n}; \quad \frac{1}{R} = \frac{1}{N} \sum\limits _{n=1}^{N}\frac{1}{R_n}; \quad J_N = \frac{1}{N}\frac{I_b}{I_c};\\
     S_N({\bf B}) = \frac{1}{N}\sum\limits _{n=1}^{N}\frac{I_{c,n}}{I_c}
    \exp\left(\frac{2\pi i}{\Phi_0} \sum\limits _{m=0}^{n-1}(<{\bf B},{\bf a}_m>)\right).
  \end{array}
  \right.
\]  
$I_{c,n}$ and ${R_n}$ are the critical current and normal resistance of the individual Josephson junctions, ${I_b}$ is the bias current, ${\Phi_0}$ is the magnetic flux quantum, and ${\bf a}_0$ is taken equal to zero.
The transfer factor $V_B=\max(\partial V/\partial B)$, where the maximum is taken within the dip, is a measure of the sensitivity of the SQIF to the applied magnetic field, and it is regulated by the SQIF geometrical parameters, such as the actual loop size distribution, the total number of loops, and the inductive interaction between them \cite{Haussler:JAP2001,Oppenlander:ASC2001}.
The uniqueness of the dip at $B=0$ and the high transfer factor $V_B$ make the SQIFs suitable for a variety of applications. In particular, SQIFs can be used as high precision sensors of absolute magnetic fields, even in noisy environments, with an expected magnetic field sensitivity of few ${\rm fT/\sqrt{Hz}}$ \cite{Oppenlander:PhysicaC-2002}. Another proposed application is a low noise multi-stage operational amplifier, in which SQIF-stages are on-chip connected in a negative-feedback configuration, resulting in an amplifier with linear and non-hysteretic characteristics\cite{Haussler:ASC2002}. 

\begin{figure}[!htbp]
  \centering
  \resizebox{\linewidth}{!}{\includegraphics*{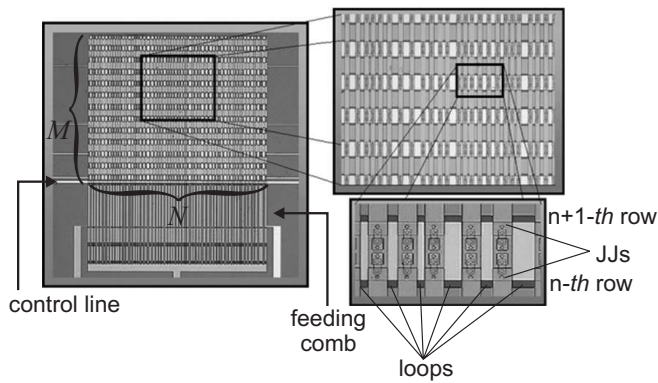}}
  \caption{
    Optical micrograph of the 2D-SQIF with $M~=~25$ and $N~=~58$. The bias current $I_b$ is supplied via a feeding comb, to reduce magnetic self-field effects. The unconventional distribution of the loop sizes is clearly visible in the enlargement.
  }
  \label{sketches}
\end{figure}

In this Letter, we present a 2D-SQIF, which consists of several parallel 1D-SQIFs connected so that they form a 2D-structure, as shown in Fig.~\ref{sketches}. The main advantage of this system is that the voltage amplitude of the dip increases with the number of rows $M$, while the dip width decreases with the number of columns $N$. In the same time, the real part of the impedance remains low ($\Re{(Z)} \propto M/N$)\cite{Oppenlander:ASC2002}. 
We systematically investigate the $V(B)$ dependences of 2D-SQIFs in a uniform magnetic field. Similarly to 1D-SQIFs, 2D-SQIFs show a unique dip at $B=0$. We have found that the voltage span of the dip scales with the number of rows, and that the transfer factor increases with the number of JJs per row. 
We have also characterized the response of the 2D-SQIF to a field gradient, artificially created by means of a control line. The experimental results may be interesting for development of 2D-SQIFs as detectors of spacial gradients of magnetic fields%
\cite{Dantsker:APL1997,Broussov:ASC2002}. 

%
In order to investigate the $V(B)$ dependences of 2D-SQIFs, we have chosen 
arrays made of 25 rows by 58 loops. In Fig.\,\ref{sketches} an optical micrograph of the device is shown. The enlargement shows the unconventional loop size distribution, characteristic of SQIF devices. In order to have bias uniformity, the bias current $I_b$ is injected in each external node of the array through bus resistors. 
To minimize self-field effects, the bias currents are applied to each column of the array in the following way. Each bias line injects current into the first row of the SQIF, as can be seen in Fig.~\ref{sketches}. The bias current of the corresponding column is extracted from the last row. Then the bias line goes down to the underlaying superconducting layer of the circuit, and returns back to the beginning of the bias comb just under the injection line. 

The voltages are measured using contact lines on the left (or right) side of the 2D-SQIF, as seen in Fig.~\ref{sketches}, across groups of 1, 5, 10, 15, 20, and 25 rows. 
The 2D-SQIFs are made of
Nb/Al-AlO$_x$/Nb Josephson tunnel junctions of area designed to be $20\,\mathrm{\mu m^2}$\footnote{HYPRES~Inc.~Elmsford~NY~10523.}.
The typical values of the critical current density are $j_c\,\approx\,1100 \,\mathrm{A/cm^2}$.
The junction capacitance is $C\,=\,600\,$pF. The junctions are overdamped (the external shunt resistance is $1 \Omega$), and the typical $I_cR$ product is about $150 \,\mathrm{\mu V}$. The McCumber parameter $\beta_c$\cite{Lih} is about 0.5.
Within a row, the loop areas range between $30 \ldots 166 \,\mathrm{\mu m^2}$. Correspondingly, the values of the parameter $\beta_L={L I_c}/{\Phi_0}$ range between $0.1 \ldots 0.5$. The total effective loop area is 
$\approx 136000\, \mathrm{\mu m^2}$ 
(1450 loops). The magnetic field is applied perpendicular to the plane of Fig.\,\ref{sketches}, either by means of an external coil or by means of an on-chip superconducting control line, placed along the first row of the 2D-SQIF (see Fig.\,\ref{sketches}). Residual magnetic fields are shielded by a cryoperm can, and low-frequency noise is filtered at room temperature by RC-filters. The experiments are carried out in liquid He ($T = 4.2\,\mathrm{K}$). 

%

\begin{figure}[!htbp]
  \centering
  \includegraphics*{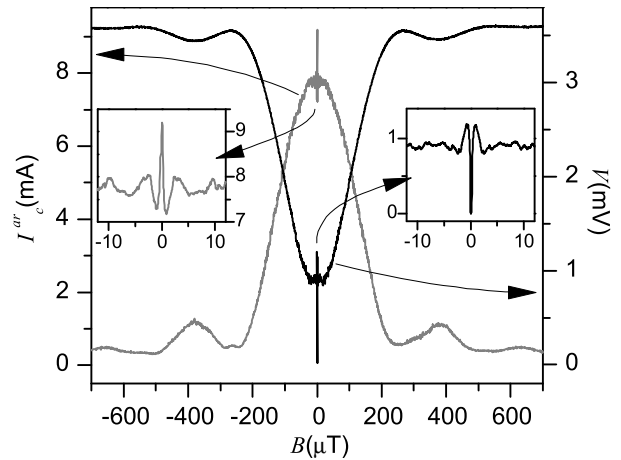}
  \caption{
    $V(B)$ (black curves, relatively to the right axes) and $I^{ar}_c(B)$ (gray curves, relatively to the left axes) dependences of the whole 2D-SQIF (25 rows) in high magnetic field, and their corresponding behavior in the low field region (insets). Both $V(B)$ dependences are measured at $I_{b}\,=\,8.1 \mathrm {mA}$. 
}
\label{large_field}
\end{figure}

We have first studied the response of the 2D-SQIF to a uniform magnetic field $B$ generated by the external coil. Figure~\ref{large_field} contains the typically measured $V(B)$ (relatively to the right axes) and $I^{ar}_c(B)$ (relatively to the left axes) dependences of the whole 2D-SQIF.  
The right inset of Fig.~\ref{large_field} shows the $V(B)$ curve of the 2D-SQIF at the $I_b=8.1\, \mathrm{mA}$ ({\it i.e.} $I_b \approx 1.1\, I^{ar}_c$), in the low magnetic field region. The unique and steep dip due to the interferometric nature of the 2D-SQIF is located at $B=0$. The voltage span of the dip $\Delta V$ is of about 1.2 mV, the side modulations are rather small and symmetric with respect to zero field. In a complementary way, the $I^{ar}_c(B)$ dependence has its maximum at $B=0$ (left inset of the figure). By further increasing the field, we do not find any periodicity in either dependence. However, the high field voltage response of the 2D-SQIF is dominated by the single junction Fraunhofer-like diffraction pattern, and the quantum interference of all cells disappears (black curve of Fig.~\ref{large_field}). A similar dependence is shown for the $I^{ar}_c(B)$ pattern (gray curve of Fig.~\ref{large_field})\footnote{As expected from the RSJ model of the single JJ, the side lobes of the $V(B)$ are less pronounced than those of the $I_c(B)$. Also, for small $I_c$, 
while the $I_c(B)$ curve shows cusp-like minima, the corresponding maxima in $V(B)$ curve have a parabolic-like behavior.}. Although the magnetic field is applied perpendicular to the loops, the screening currents generated in the asymmetric top and bottom electrodes of the junctions (see enlarged details of Fig.~\ref{sketches}) might induce a nonzero magnetic field in the plane of the JJs, and therefore cause the single junction response.

\begin{figure}[!htbp]
  \centering
  \includegraphics*{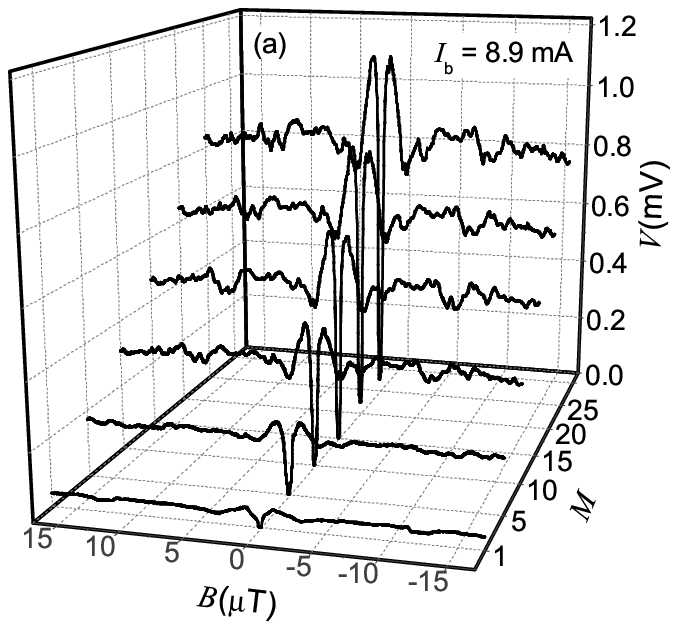}\\
  \includegraphics*{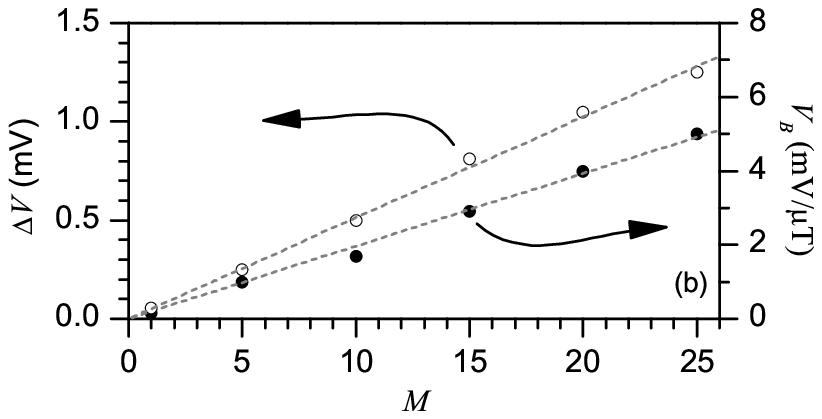}
  \caption{
    (a) $V(B)$ dependences of various groups of rows $M$ of the 2D-SQIF. The rows are simultaneously biased at $I_b=8.9\, \mathrm {mA}$, and the voltages are measured across groups of $M=\,$1, 5, 10, 15, 20, and 25 rows. (b) Dependences of the voltage span $\Delta V$ (open symbols) and of the transfer factor $V_B$ (closed symbols) on $M$, and their linear 
    fits (dashed lines). 
  }
  \label{scaling}
\end{figure}

Figure~\ref{scaling}(a) shows the $V(B)$ curves measured across groups of $M=$ 1, 5, 10, 15, 20 and 25 rows, while the whole 2D-SQIF is biased ($I_b=8.9 \,\mathrm{mA}$). $\Delta V$ is proportional to the number of rows $M$. 
Thus, within the 2D-SQIF, individual rows behave as insulated 1D-SQIFs, and operate simultaneously so that each row contributes by the same amount to the total voltage. In Fig.~\ref{scaling}(a), all reported $V(B)$ curves present the dip at the same value of magnetic field ($B~=~0$), and among the various groups of rows the width of the dip $\Delta B$ is almost the same. This result confirms the theoretical predictions that $\Delta B$ is indeed only a function of the number of JJs per row\cite{Oppenlander:PhysicaC-2002}. In the same time, equal dip widths are a sign of magnetic field uniformity from row to row of the 2D-SQIF. In the case when individual rows of a given group would show the dip at slightly different values of $B$, the measured $V(B)$ dependence for that group is expected to exhibit a larger $\Delta B$. In addition, we have observed that the profile of the side modulations (outside the dip) is similar for all groups, which is interpreted in terms of field uniformity, as well as uniformity of the junction parameters and of the bias current. 
To emphasize the linear increase of dip height with the row number, in Fig.~\ref{scaling}(b) reported are the measured $\Delta V$ vs. $M$ (relatively to the left axes). The maximum voltage span of the dip measured across 25 rows is about 1.25 mV. The linear fit of the data gives $\Delta V=51\, \mathrm {\mu V}$ per row. Relatively to the right axes, reported are the values of the corresponding transfer factor $V_B$. The maximum $V_B$ measured for the 25 rows is about $V_B=5\, \mathrm {mV}/{\mu T}$. In this case, the linear fit of the data gives $V_B=197\, \mathrm {\mu V}/{\mu T}$ per row.

\begin{figure}[!htbp]
  \centering
  \includegraphics{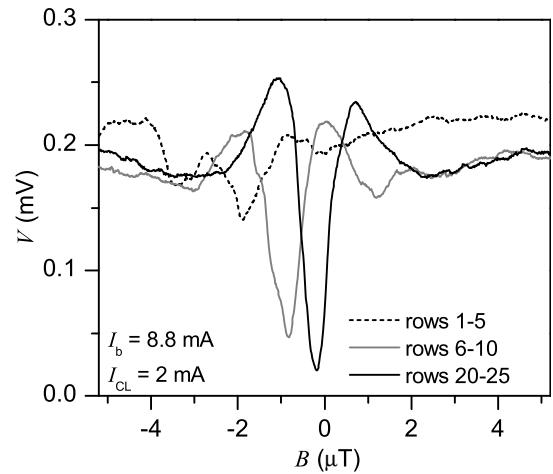}
  \caption{
   $V(B)$ dependences of three neighboring groups of five rows each, in a field gradient generated by $I_{CL}=2\, \mathrm {mA}$. The response of individual groups to the magnetic field of the coil $B$ depends on their distance from the control line. 
 }
 \label{gradient}
\end{figure}

When passing a current $I_{CL}$ through the control line, a magnetic field gradient is generated in the 2D-SQIF. The field decays as $1/r$, where $r$ is the distance from the control line. The 2D-SQIF exhibits a sensitivity to the field gradient. We have measured the $V(B)$ dependences of the various groups of rows at different constant values of $I_{CL}$, while sweeping the coil field $B$. Increasing $I_{CL}$, the minimum of the dip is progressively shifted from $B=0$, 
as a different coil field $B$ is required to compensate the field gradient. 
Figure~\ref{gradient} shows, as an example, the $V(B)$ dependences of three groups of five rows [rows 1--5 (dashed line), rows 6--10 (continuous gray line), and rows 20--25 (continuous black line)] at $I_{CL}=2\,\mathrm {mA}$. As one can see, the $V(B)$ dependence of the group of rows nearest to the control line (rows 1--5) shows the largest deviation from $B=0$. In this case, a compensation field of about $-2\,\mathrm {\mu T}$ is necessary to have the dip. In the same time, the voltage span of the dip is significantly reduced. Since this group of rows is located where there is the largest field gradient, each individual row requires a slightly different compensation field. Thus, the series measured $V(B)$ dependence of this group shows a dip with reduced span and, in the same time, larger width.        
These effects are less evident in the neighboring group (rows 6--10). For the farthest group (rows 20--25), the presence of the field gradient is almost unnoticeable.

In conclusion, the extension of the properties of Superconducting Quantum Interference Filters to 2D-SQIFs has been successfully realized. As in the 1D case, also 2D-SQIFs have a non-periodic voltage response to the applied magnetic field, with a unique dip around $B=0$. The voltage span of the dip and the transfer factor scale with the number of rows operating at the same working point. The 2D-SQIFs are sensitive to a field gradient generated by a control line and superimposed to the homogeneous field coil: depending on the value of $I_{CL}$, the $V(B)$ curves of neighboring groups of rows show a degradation of the dip and a shift of the minimum from $B=0$. The experimental results may be interesting for development of 2D-SQIF based sensors for mapping spatial gradients of magnetic fields. 

\bibliography{references}   
\bibliographystyle{apsrev}
\end{document}